\documentclass[english,aps,prl,twocolumn,superscriptaddress,showpacs,amsmath]{revtex4-2}

\usepackage[T1]{fontenc}
\usepackage[latin9]{inputenc}
\usepackage{amssymb}
\usepackage{bm}
\usepackage{amsmath}
\usepackage{mathtools}
\usepackage{graphicx}
\usepackage{babel}
\usepackage{xcolor}
\usepackage{units}
\usepackage{esint}
\usepackage{natbib}
\usepackage{caption}
\usepackage[breaklinks=true,colorlinks=true,urlcolor=blue,
citecolor=blue,linkcolor=blue,bookmarks=false]{hyperref}
\usepackage{lineno}
\setcitestyle{journalcolor = blue}

\begin{document}

\title{Universality in quantum critical flow of charge and heat in ultra-clean graphene}

\author{Aniket Majumdar}
\email{aniketm@iisc.ac.in}
\affiliation{Department of Physics, Indian Institute of Science, Bangalore 560012, India}
\author{Nisarg Chadha}
\affiliation{Department of Physics, Indian Institute of Science, Bangalore 560012, India}
\affiliation{Department of Physics, Harvard University, Cambridge, MA 02138, USA}
\author{Pritam Pal}
\affiliation{Department of Physics, Indian Institute of Science, Bangalore 560012, India}
\author{Akash Gugnani}
\affiliation{Department of Physics, Indian Institute of Science, Bangalore 560012, India}
\author{Bhaskar Ghawri}
\affiliation{Department of Physics, Indian Institute of Science, Bangalore 560012, India}
\author{Kenji Watanabe}
\affiliation{Research Center for Electronic and Optical Materials, National Institute for Materials Science, 1-1 Namiki, Tsukuba 305-0044, Japan}
\author{Takashi Taniguchi}
\affiliation{Research Center for Materials Nanoarchitectonics, National Institute for Materials Science, 1-1 Namiki, Tsukuba 305-0044, Japan}
\author{Subroto Mukerjee}
\email{smukerjee@iisc.ac.in}
\affiliation{Department of Physics, Indian Institute of Science, Bangalore 560012, India}
\author{Arindam Ghosh}
\email{arindam@iisc.ac.in}
\affiliation{Department of Physics, Indian Institute of Science, Bangalore 560012, India}
\affiliation{Center for Nano Science and Engineering, Indian Institute of Science, Bangalore 560012, India}

\maketitle

\textbf{Close to the Dirac point, graphene is expected to exist in quantum critical Dirac fluid state, where the flow of both charge and heat can be described with a dc electrical conductivity $\sigma_\mathrm{Q}$, and thermodynamic variables such as the entropy and enthalpy densities \cite{son2007quantum,hartnoll2007theory}. Although the fluid-like viscous flow of charge is frequently reported in state-of-the-art graphene devices \cite{crossno2016observation,bandurin2016negative,ku2020imaging,bandurin2018fluidity,krishna2017superballistic,xin2023giant,block2021observation,ghahari2016enhanced,gallagher2019quantum, kumar2022imaging, sulpizio2019visualizing,huang2023electronic, tan2022dissipation}, the value of $\sigma_\mathrm{Q}$, predicted to be quantized and determined only by the universality class of the critical point, has not been established experimentally so far~\cite{hartnoll2007theory,damle1997critical}. Here we have discerned the quantum critical universality in graphene transport by combining the electrical ($\sigma$) and thermal ($\kappa_\mathrm{e}$) conductivities in very high-quality devices close to the Dirac point. We find that $\sigma$ and $\kappa_\mathrm{e}$ are inversely related, as expected from relativistic hydrodynamics, and $\sigma_\mathrm{Q}$ converges to $\approx (4\pm 1)\times e^2/h$ for multiple devices, where $e$ and $h$ are the electronic charge and the Planck's constant, respectively. We also observe, (1) a giant violation of the Wiedemann-Franz law where the effective Lorentz number exceeds the semiclassical value by more than 200 times close to the Dirac point at low temperatures, and (2) the effective dynamic viscosity ($\eta_\mathrm{th}$) in the thermal regime approaches the holographic limit $\eta_\mathrm{th}/s_\mathrm{th} \rightarrow \hbar/4\pi k_\mathrm{B}$ within a factor of four in the cleanest devices close to the room temperature, where $s_\mathrm{th}$ and $k_\mathrm{B}$ are the thermal entropy density and the Boltzmann constant, respectively. Our experiment addresses the missing piece in the potential of high-quality graphene as a testing bed for some of the unifying concepts in physics.}

\begin{figure*}[tbh]
	\includegraphics[width=1.0\textwidth]{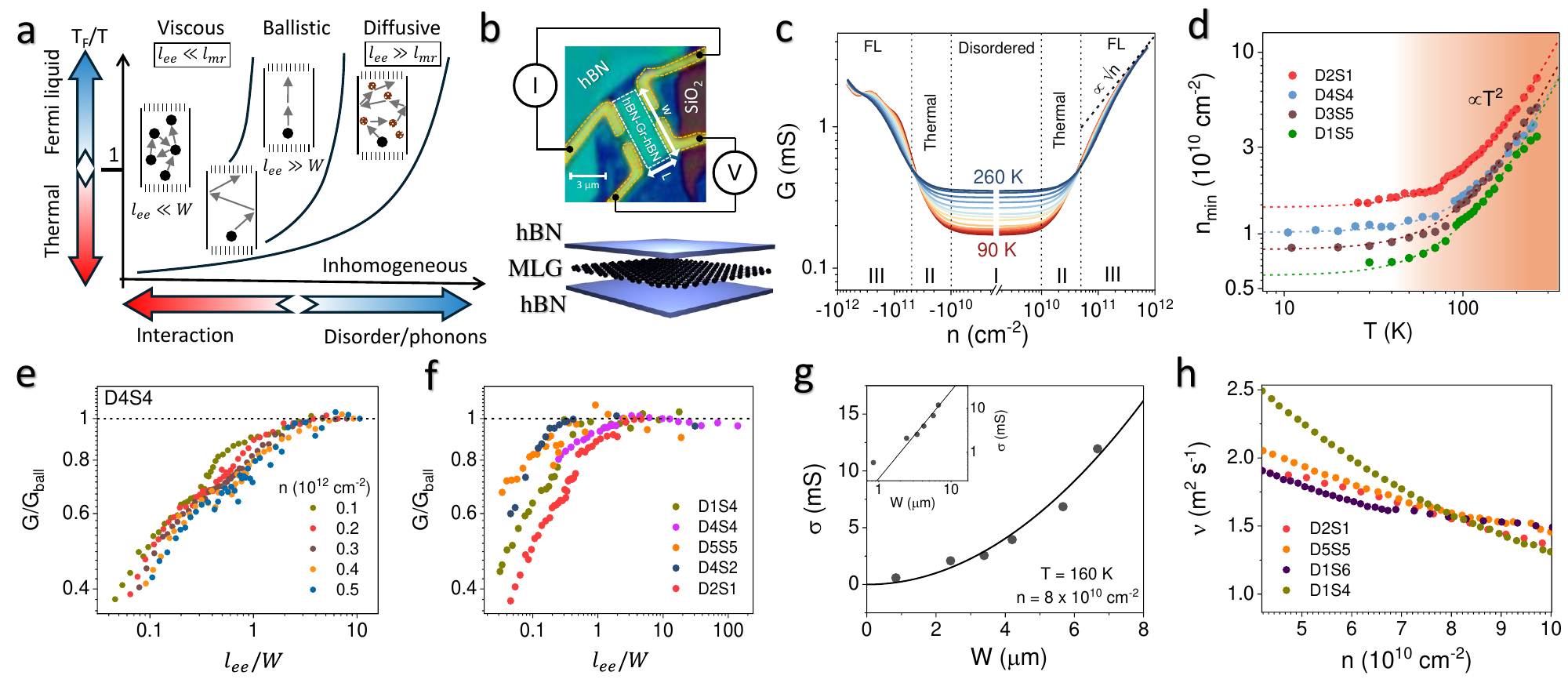}
	%\makebox[\textwidth][l]{\includegraphics[width=0.48\textwidth]{Fig_1.pdf}}
    \caption{\textbf{Viscous electron flow in ultra-clean graphene devices:} (a) Different regimes of electron transport in graphene, based on the interplay between momentum-conserving and momentum-relaxing collisions. (b) (Top) Optical micrograph of the device along with the circuit diagram used for conducting the electrical transport measurements. (Bottom) Schematic of the heterostructure. (c) Electrical conductance of device D4S4 as a function of the carrier density $n$ for $90$~K $\leq T \leq 260$~K. The dashed line labelled $G_\mathrm{ball}$ indicates the ballistic conductance for the measured channel. (d) Charge inhomogeneity ($n_\mathrm{min}$) as a function of $T$ for five different devices. The colour gradient in the background indicates the transition from a disorder-driven regime to a thermally-driven transport regime. (e) Electrical conductance (normalised by the ballistic conductance, $G_\mathrm{ball}$) of D4S4 as a function of $l_\mathrm{ee}/W$ for different carrier densities, varying from $n = 1 \times 10^{11}$~cm$^{-2}$ to $5 \times 10^{11}$~cm$^{-2}$. (f) Normalised conductance as a function of the Knudsen number for five different devices. (g) Electrical conductivity as a function of width $W$ at $n = 8 \times 10^{10}$~cm$^{-2}$ and $T = 160$~K. For the chosen values of $n$ and $T$, we have observed that $l_\mathrm{ee}/W \leq 0.5$. The solid line scales as $W^2$ and serves as a guide to the eye. Inset shows the same dataset in double logarithmic scale where the straight line represents quadratic dependence. (h) Kinematic viscosity $\nu$ as a function of $n$ for four different devices, measured at $T = 180$~K. }
    \label{figure_1}
\end{figure*}

Transport in real graphene devices is determined by two competing length scales: The electron-electron scattering length $l_\mathrm{ee}$ and the momentum relaxation length $l_\mathrm{mr}$, both of which depend on the carrier density $n$ (or the Fermi Temperature $T_\mathrm{F}$ $= \hbar v_F \sqrt{\pi n}$, $v_F$ being the Fermi velocity) and the temperature $T$ (see schematic in Fig.~\ref{figure_1}a). In the diffusive and inhomogeneous regimes, scattering from, for example, the Coulomb impurities or phonons, do not conserve the momentum, resulting in $l_\mathrm{mr}\ll l_\mathrm{ee}$. As the disorder decreases, the channel can be ballistic when the device dimension is $\ll l_\mathrm{ee}$,~$l_\mathrm{mr}$ or viscous, when $l_\mathrm{ee}$ becomes the shortest length scale. The viscous flow is described well by the electronic analogue of the Navier-Stokes equation \cite{levitov2016electron} when $T_\mathrm{F}/T \gg 1$, and manifests in parabolic propagation profile \cite{sulpizio2019visualizing,ku2020imaging}, negative non-local resistance \cite{bandurin2016negative} or super-ballistic electrical conduction \cite{krishna2017superballistic}, among others. 

In the opposite (thermal) limit of $T_\mathrm{F}/T \ll 1$, nearly unscreened Coulomb interaction and linear dispersion tune to a Lorentz-invariant quantum critical point \cite{hartnoll2007theory, fritz2008quantum}, where the current relaxes by collisions among the thermally excited electrons and holes, %at the maximum (Planckian) rate allowed by the uncertainty principle, $\tau_\mathrm{p}^{-1} \approx \alpha^2k_\mathrm{B}T/\hbar$, $\alpha$ being the fine structure constant. 
while the energy propagates unimpeded due to conservation of momentum during such collisions. This naturally violates the Wiedemann-Franz (WF) law. For relativistic hydrodynamics, quantum critical $\sigma$ and $\kappa_\mathrm{e}$ are predicted to be \cite{lucas2016transport, xie2016transport, lucas2018hydrodynamics},  

\begin{equation}
\label{sigma}
    \sigma(n) = \sigma_\mathrm{Q} + \frac{e^2v_\mathrm{F}n^2l_\mathrm{mr}}{\mathcal{H}}
\end{equation}
\noindent and
\begin{equation}
\label{kappa}
    \kappa_\mathrm{e}(n) = \frac{v_\mathrm{F}l_\mathrm{mr}{\cal H}\sigma_\mathrm{Q}}{T\sigma(n)}
\end{equation}

\noindent where, ${\cal H}$ and $v_\mathrm{F}$ are the enthalpy density and Fermi velocity, respectively. So far, several experimental reports on the breakdown of the WF law \cite{crossno2016observation}, enhanced thermoelectric power \cite{ghahari2016enhanced}, giant quasi-linear magnetoresistance \cite{xin2023giant}, near-Planckian scattering rates and efficient thermal diffusion inferred from the terahertz \cite{gallagher2019quantum} and pump-probe \cite{block2021observation} spectroscopic techniques, provided valuable insight to the formation of a Dirac fluid, but unambiguous experimental evaluation of $\sigma_\mathrm{Q}$ has not been possible. This is because the measured $\sigma$ for $n \rightarrow 0$ yields non-universal results, limited by device-dependent spatial inhomogeneity, or puddles, of charge close to the Dirac point \cite{adam2007self} (see Supplementary Section S1). In the absence of this, the quantum critical behaviour itself remains unsettled, especially in the presence of alternative models that propose violation of the WF law in marginally gapped graphene within the Fermi liquid framework \cite{tu2023wiedemann1}.

In this work, we have combined dc charge and heat transport to explore the quantum critical conductivity in graphene. We measured both $\sigma$ and $\kappa_\mathrm{e}$ in multiple graphene devices with low charge inhomogeneity $\simeq 5 - 10\times10^{9}$~cm$^{-2}$, where the momentum relaxation, particularly at low temperatures, occurs at the boundary in most of the devices \cite{xin2023giant}. Combining $\sigma$ and $\kappa_\mathrm{e}$ as in Eqn.~\ref{kappa}, and assigning $l_\mathrm{mr} = \mathrm{min}(W,l_\mathrm{mfp})$, where $W$ and $l_\mathrm{mfp} = h\sigma/2e^2k_\mathrm{F}$ are the geometric channel width and the mean free path of carriers in the channel, respectively, we obtain a device-insensitive estimate of $\sigma_\mathrm{Q} \simeq (4\pm1)\times e^2/h$, where the error represents variation across multiple devices. Fundamentally, this {\it all-experimental} strategy to determine $\sigma_\mathrm{Q}$ depends on the high quality of our devices which makes the thermal limit ($T_\mathrm{F}/T < 1$) accessible over a broad range of $T$ for $n \gtrsim n_\mathrm{min}(0)$, {\it i.e.}, when $n$ is well-defined.

\section{Electron viscosity in the ultra-clean graphene devices}

All devices are single layers of graphene encapsulated with hexagonal boron nitride (hBN), which were etched into small rectangles of length $L$ ($\approx 1$ to 2~$\mu$m) and width $W$, where $W$ was varied between $\approx 1$ to 8~$\mu$m (Figs.~\ref{figure_1}b, details in Methods, Table 1 and Supplementary Section S2). The variation of conductance ($G$) in a typical device (D4S4, see Table 1) with $n$ for $T = 90 - 260$~K is shown in Fig.~\ref{figure_1}c. In the low-$n$ limit, $G$ becomes insensitive to $n$ below a characteristic scale $n_\mathrm{min}(T)$, which increases with increasing $T$. The $T$-dependence of $n_\mathrm{min}(T)$, as shown for multiple devices in Fig.~\ref{figure_1}d, can be described with $n_\mathrm{min}(T) = n_\mathrm{min}(0)+\beta n_\mathrm{th}(T)$, where $n_\mathrm{min}(0)$ and $n_\mathrm{th}(T)=(2\pi^3/3)(k_\mathrm{B}T/hv_\mathrm{F})^2$ are the intrinsic inhomogeneity and the thermally excited carrier density, respectively (the prefactor $\beta$ is $\approx 0.7 - 0.8$ for most devices, likely due to the residual charge inhomogeneity \cite{ponomarenko2024extreme}). For the present experiments, we chose devices with $n_\mathrm{min}(0) \lesssim 10^{10}$~cm$^{-2}$, which usually exhibited lower Raman linewidth and were occasionally subjected to post-fabrication current annealing (see Methods and Supplementary Section S2). The best devices exhibited carrier mobility as high as $5\times10^5$~cm$^2$/V.s  and $6\times10^6$~cm$^2$/V.s at room temperature and low $T$ ($\approx 10$~K), respectively, for $n\approx 10^{11}$~cm$^{-2}$ (see Supplementary Section S3 and Table I).

Fig.~\ref{figure_1}c identifies three distinct ranges in $n$. I: $n < n_\mathrm{min}(0)$, where the system is spatially inhomogeneous; II: the quasi-thermal regime, where $n_\mathrm{min}(0) < n \lesssim 8\times10^{10}$~cm$^{-2}$, for which $T_\mathrm{F}$ is lower than the maximum experimental $T$ ($300$~K), and, III: $n \gtrsim 10^{11}$~cm$^{-2}$, where we expect the system to behave as Fermi liquid at all experimental $T$. In the large $n$ ($\gtrsim 5\times10^{11}$~cm$^{-2}$) limit of regime III, the (four-terminal) conductance $G$ approaches $\sim \sqrt{n}$ with a nearly $T$-independent magnitude $G_\mathrm{ball}$ (dashed line in Fig.~\ref{figure_1}c), that lies within a factor of about two of the Landauer-Sharvin conductance ($G_\mathrm{ls} = (4e^2/\pi h)k_\mathrm{F}W$). This is the ballistic regime where the Landauer-Sharvin conductance is delocalised into the bulk as a result of hydrodynamic flow that mediates the transfer of momentum among the propagating modes \cite{kumar2022imaging}. $G/G_\mathrm{ball}$ reduces from unity as $n$ is decreased which may signify the onset of viscous flow (Fig.~\ref{figure_1}a). To examine this, we plot $G/G_\mathrm{ball}$ as a function of $l_\mathrm{ee}/W$ for a broad range of $n$ in regime III of device D4S4 for which $l_\mathrm{mr} \approx W$ (Fig.~\ref{figure_1}e). We evaluated $l_\mathrm{ee} \approx \hbar v_\mathrm{F}T_\mathrm{F}/\alpha^2k_\mathrm{B}T^2$ using the effective fine structure constant $\alpha \approx 0.5$ \cite{muller2008quantum}. The onset of decrease in $G/G_\mathrm{ball}$ at $l_\mathrm{ee}/W \approx 1$ is evident in Fig.~\ref{figure_1}e. Since $l_\mathrm{ee}/W$ effectively represents the Knudsen number $\zeta (= l_\mathrm{ee}/l_\mathrm{mr})$ \cite{bandurin2018fluidity}, this constitutes a strong evidence of the onset of Poiseuille flow. This behaviour was found to be generic in all high mobility devices, as shown for five different devices at $n = 10^{11}$~cm$^{-2}$ in Fig.~\ref{figure_1}f, for which the momentum relaxation is expected to occur mainly at the boundaries (see Supplementary Section S3 for further details). 

Strong Poiseuille-like flow of degenerate electrons requires the electronic Gurzhi length $D_\mathrm{\nu} = \sqrt{\nu l_\mathrm{mr}/v_\mathrm{F}}$ to exceed $W$, where the conductivity $\sigma = GL/W = e^2n^2W^2/12\eta$ (Supplementary Sections S5 and S6) is then expected to vary quadratically with $W$ \cite{li2022hydrodynamic} (here $\eta$, $\nu=\eta/\rho_\mathrm{m}$, and $\rho_\mathrm{m}$ are the dynamic viscosity, the kinematic viscosity, and the areal density of effective mass, respectively). In Fig.~\ref{figure_1}g, $\sigma$ in multiple devices at fixed $n$ ($\approx 8\times10^{10}$~cm$^{-2}$) and $T$ ($=160$~K), such that $T_\mathrm{F}\gtrsim T$ and $l_\mathrm{ee}/W <1$ are simultaneously satisfied, is consistent with a quadratic dependence on $W$ (see Fig.~\ref{figure_1}g inset). The quantitative estimate of $\nu$, obtained from the $n$-dependence of $\sigma$ at 180~K, is shown in Fig.~\ref{figure_1}h, where its magnitude $\sim 1.5 - 2$~m$^2$/s is nearly device-independent and agrees with previous reports \cite{bandurin2016negative, krishna2017superballistic}, as well as the theoretical calculations \cite{pellegrino2016electron}. Since $l_\mathrm{mr}$ is limited by the boundaries in these devices, the observed $\nu$ confirms $D_\nu \gtrsim W$ and hence the strong Poiseuille-like flow. The differential resistance $dV/dI$ also exhibits the characteristic non-monotonic dependence on the drive current density (Extended data Fig.~1), where the initial increase, followed by decrease, in $dV/dI$ is attributed to crossover from the quasi-ballistic Knudsen flow to the Poiseuille regime.

\begin{figure*}[tbh]
	\includegraphics[width=1.0\textwidth]{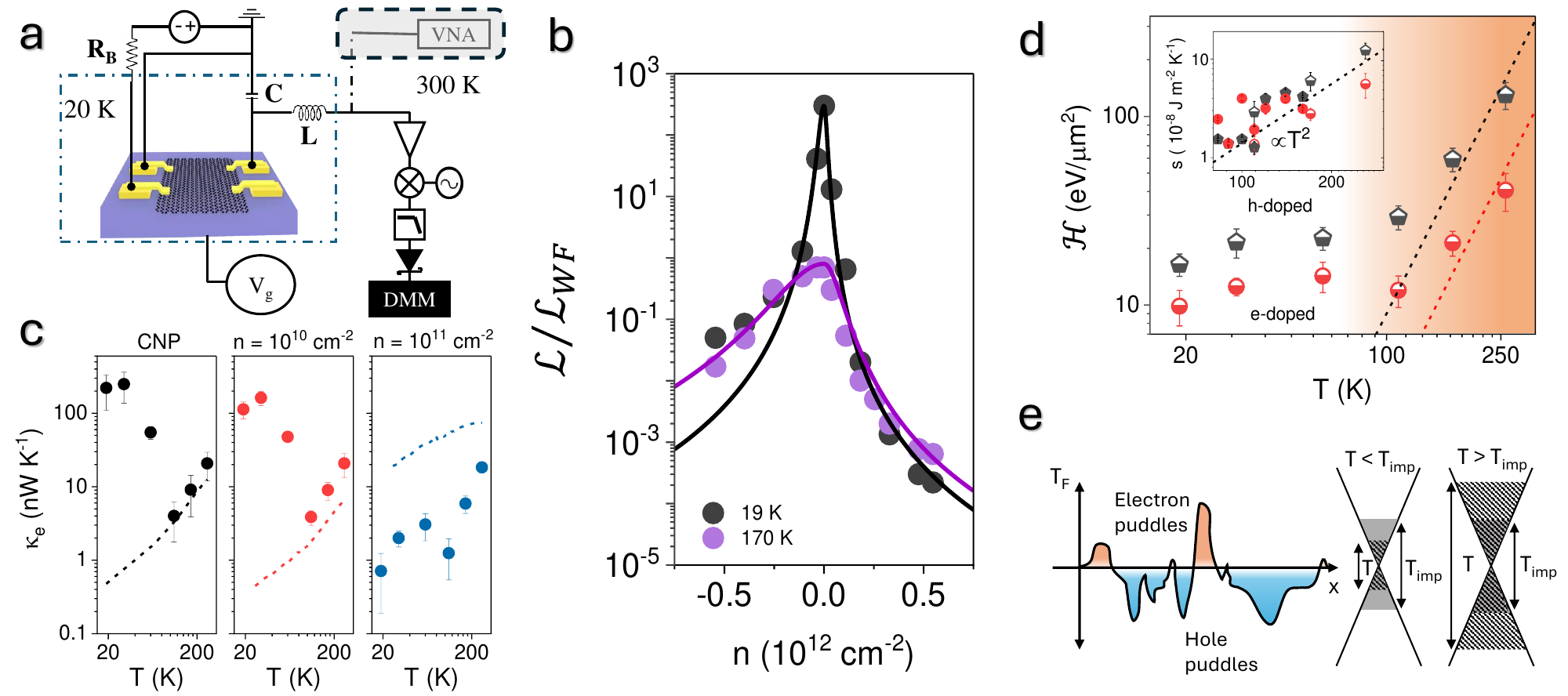}
	%\makebox[\textwidth][l]{\includegraphics[width=0.48\textwidth]{Fig_2.pdf}}
	\caption{ \textbf{ Violation of the WF Law for hydrodynamic electrons:} (a) Circuit diagram for measuring Johnson-Nyquist noise of hot electrons in our graphene devices, when subjected to Joule heating by in-plane electric fields. The LC network acts as a tank circuit for impedance matching the graphene device to the $50$~$\Omega$ noise measurement circuit. (b) Normalised Lorentz number for device D2S1 as a function of $n$ for $T = 19$~K and $170$~K. The solid lines are theoretical fits of the experimental data, as per Eqn.~\ref{L}. The electron-doped and hole-doped data points have been fitted independently using different values for the fitting parameters. (c) Electronic component of thermal conductivity for device D2S1 as a function of $T$ for 3 different number densities, from the charge neutrality point to a highly electron-doped regime. The dashed lines indicate the variation of thermal conductivity with $T$, assuming the normalised Lorentz number ${\cal L}/{\cal L}_\mathrm{WF} = 1$. (d) Enthalpy density ($\cal{H}$) for D2S1 as a function of $T$. The dashed lines (black - hole doped, red - electron doped) highlight $T^3$-like asymptotic behaviour. The colour gradient in the background signifies a transition from a disorder-driven regime to a thermally-driven transport regime. [Inset] Entropy density for D2S1 as a function of $T$ for $T > 80$~K. The black dashed line is the $T^2$ fit of the experimentally obtained data and is within a factor of 2 of what is expected from the theoretical expression in Ref.~\cite{yudhistira2023non}. (e) Schematic showing the interplay between two different types of electrical transport - disorder-driven and thermal excitation-driven. For $T<T_\mathrm{imp}$, the thermal energy of the impurities is greater than the thermal energy of electrons and hence the electrical transport is dominated by charged impurities and defects, whereas for $T>T_\mathrm{imp}$, the electronic thermal energy dominates and electron-electron interactions drive electrical transport in this regime.}
	\label{figure_2}
\end{figure*}

\section{Thermal conductivity and the Wiedemann-Franz law}
To examine the current and energy relaxation pathways, we have used Johnson noise thermometry \cite{betz2012hot} to determine the effective Lorentz number ${\cal L} = \kappa_\mathrm{e}/\sigma T$, and thereby the electronic thermal conductivity ($\kappa_\mathrm{e}$), at varying $n$ and $T$. The experimental strategy, schematically shown in Fig.~\ref{figure_2}a (details in Supplementary Section S4), involves Joule heating the graphene layer with an external electric field $E$ and evaluating the resulting increase in the electron temperature ($T_\mathrm{e}$) from the Johnson-Nyquist noise magnitude \cite{crossno2016observation}. The weak coupling of electrons and phonons in graphene~\cite{ferrari2006raman, fong2013measurement} makes the Joule dissipation $P_\mathrm{J} = \sigma E^2$ rather effective to heat the electrons, leading to $T_\mathrm{e} \approx LE/\sqrt{\cal L}$ at large $E$, as shown for sample D2S1 for different $n$ (at $T = 30$~K) and different $T$ (at $n = 5\times10^{11}$~cm$^{-2}$) in Extended Data Fig.~2a and 2b, respectively. The dashed lines are fits according to Eqn.~\ref{Te} (Methods) obtained from the spatially averaged solution for $T_\mathrm{e}$ from the heat diffusion equation with ${\cal L}$ as the (only) fit parameter. Evidently, ${\cal L}$ is strongly $n$ and $T$-dependent, which underlines the breakdown of the WF law. The magnitude of ${\cal L}$ in the electron-doped part of D2S1 varies over six orders of magnitude as a function of $n$ at low $T$ (Fig.~\ref{figure_2}b, Extended Data Fig.~3), with a maximum value (${\cal L}_0$) at the Dirac point that can exceed the universal magnitude ${\cal L}_\mathrm{WF} = (\pi^2/3)^2(k_\mathrm{B}/e)^2 = 2.44\times10^{-8}$~V$^2$K$^{-2}$ from the WF law by nearly 300 times. Correspondingly, $\kappa_\mathrm{e} = {\cal L}\sigma T$ close to the Dirac point also exceeds the WF magnitude by similar order at low $T$. At large $n$ however, {\it e.g.} at $n=10^{11}$~cm$^{-2}$, $\kappa_\mathrm{e}$ is much lower than that expected from the WF law at all $T$, even though the electrical conductivity is significantly higher (Fig.~\ref{figure_2}b). The behaviour remains similar, albeit reduced, in lower mobility devices (D1S1, Extended Data Fig.~4) and indicates decoupling of the charge and heat flow expected in the hydrodynamic regime where ${\cal L}(n,T)$ varies as~\cite{li2022hydrodynamic, lucas2018hydrodynamics},

\begin{equation}
    {\cal L}(n,T) = \frac{1}{e^2}{{\left [\frac{s(T)n_0(T)}{n^2+n^2_0(T)}\right ]}}^2
\label{L}
\end{equation}

\noindent Here, $n_0(T)$ is an effective density scale that determines the intrinsic conductivity of the electron fluid \cite{li2022hydrodynamic} (Extended data Fig.~5a). Semi-classical analysis considering small band gap opening at the Dirac point and various scattering mechanisms fail to explain the observed behaviour (Supplementary Section S7).
Eqn.~\ref{L}, however, fits very well (solid lines in Fig.~\ref{figure_2}c) to the experimentally observed dependence of ${\cal L}$ on $n$, especially in the electron doped regime. These fits also provide experimental estimation of two key thermodynamic variables associated with the hydrodynamic flow, namely the entropy $s(T) = en_0\sqrt{{\cal L}_0}$ and, from Gibb's equation, the enthalpy ${\cal H}(T) = Ts(T)$ densities. The dependence of ${\cal H}$ and $s$ on $T$ are shown in Fig.~\ref{figure_2}d and its inset, respectively. 

An important temperature scale, identified by the non-monotonic behaviour of ${\cal L}_0$ (Extended data Fig.~5b), is $T \sim T_\mathrm{imp} = \hbar v_\mathrm{F}\sqrt{\pi n_\mathrm{min}(0)}/2k_\mathrm{B}\approx 80$~K, determined by the fluctuations in local chemical potential from intrinsic disorder.
For $T \ll T_\mathrm{imp}$, the incompressibility of graphene arrests the fluctuations in $n$ to $\sim n_\mathrm{min}(0)$, which saturates ${\cal H}$ to the $T$-independent Fermi liquid contribution (${\cal H}_0$). The thermal component ${\cal H}_\mathrm{th} \sim T^3$ dominates only for $T\gtrsim T_\mathrm{imp}$ (dashed lines in Fig.~\ref{figure_2}d). The scenario is schematically explained in Fig.~\ref{figure_2}e. For D2S1, we find ${\cal H}_0 \sim \rho_\mathrm{m0}v_\mathrm{F}^2 = 3.5$~eV/$\mu$m$^2$ within a factor of $\sim 3$ of the observed enthalpy at low temperatures (here, $\rho_\mathrm{m0}$ is the mass density at $n_\mathrm{min}(0)$). Moreover, an estimate of ${\cal L}_0 \approx v_\mathrm{F}W{\cal H}_0/\sigma_\mathrm{min}T^2$ at $T \ll T_\mathrm{imp}$ from Eqn.~\ref{kappa} suggests ${\cal L}_0/{\cal L}_\mathrm{WF} \approx 500$ at $T = 20$~K, which is within a factor of two of that observed experimentally.

\begin{figure*}
	\includegraphics[width=1.0\textwidth]{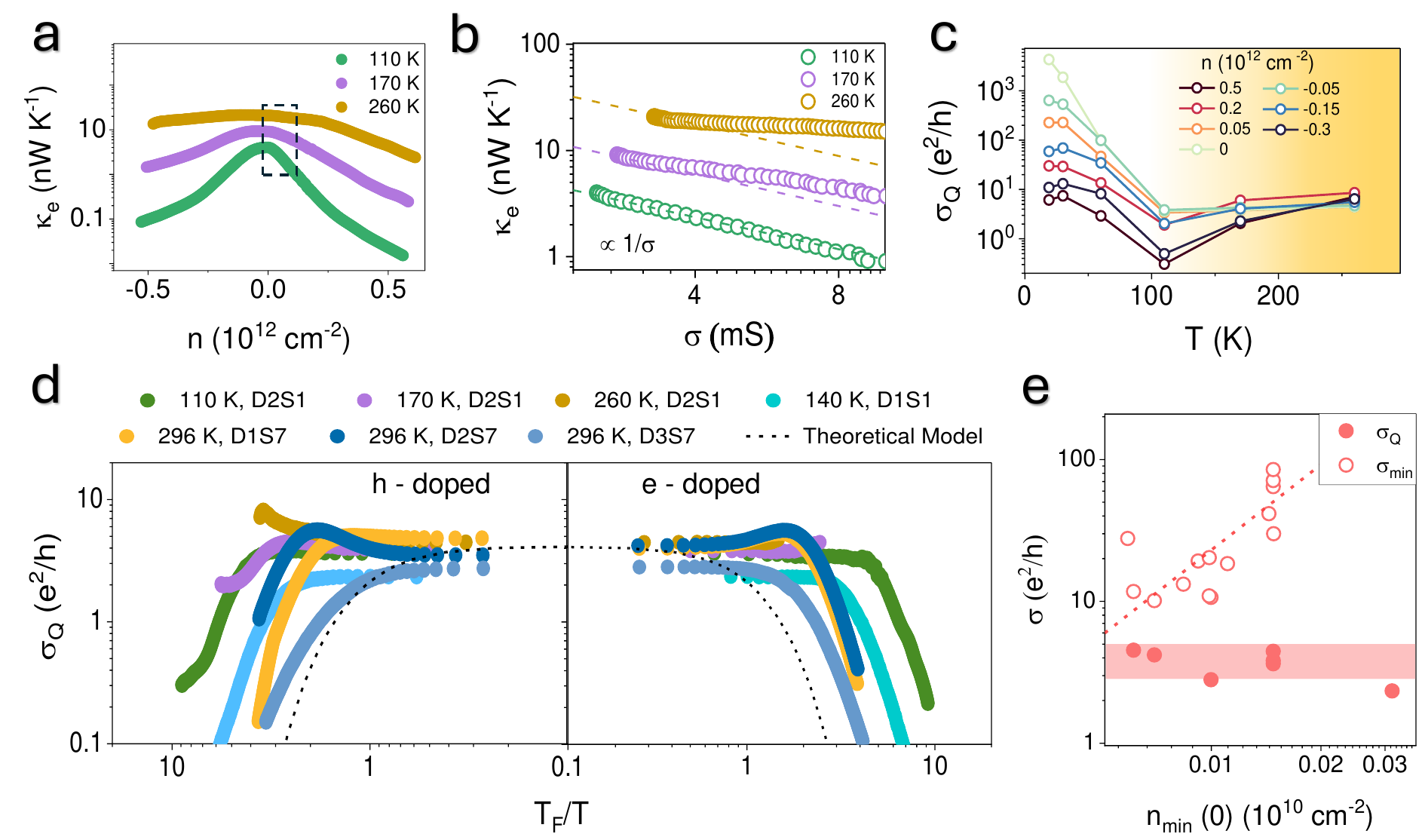}
	%\makebox[\textwidth][l]{\includegraphics[width=1.0\textwidth]{Fig_3.pdf}}
	\caption{ \textbf{Universality of the quantum critical conductivity:} (a)  Electronic component of thermal conductivity ($\kappa_\mathrm{e}$) for device D2S1 as a function of $n$ for $T = 110$~K, $170$~K and $260$~K. (b) $\kappa_\mathrm{e}$ as a function of electrical conductivity $\sigma$ for device D2S1 at three different temperatures from the region indicated by the bounding box in panel (a). The dashed lines indicate a $1/\sigma$ dependence and serve as a guide to the eye. (c) $\sigma_\mathrm{Q}$ for D2S1 as a function of $T$ for a range of densities from $n = -3 \times 10^{11}$~cm$^{-2}$ (hole-doped) to $n = 5 \times 10^{11}$~cm$^{-2}$ (electron-doped). The shaded region corresponds to temperatures greater than the scale of background chemical potential fluctuations. (d) $\sigma_\mathrm{Q}$ as a function of the ratio of Fermi temperature to absolute temperature ($T_\mathrm{F}/T$), calculated at different temperatures for four different devices. The dashed line is based on theoretical calculations performed in Ref.~\cite{muller2008quantum}. (e) Comparison of the quantum critical conductivity $\sigma_\mathrm{Q}$ (obtained from thermal transport measurements) and the minimum electrical conductivity $\sigma_\mathrm{min}$ (obtained from electrical transport measurements at $T<60$~K), as a function of the intrinsic charge inhomogeneity $n_\mathrm{min}(0)$. The dashed line scales as $n^2_\mathrm{min}(0)$ and serves as a guide to the eye.}
	\label{figure_3}
\end{figure*}

\section{Quantum critical transport in the thermal regime}
Having established the dominance of the electron-electron scattering rate over that of momentum relaxation in our devices, we now focus on the thermal regime where we expect the system to behave as a quantum-critical Dirac fluid. The rapid increase in ${\cal H}(T)$ at $T\gtrsim T_\mathrm{imp}$ (Fig.~\ref{figure_2}d) suggests the onset of the thermal regime, where the corresponding entropy density $s_\mathrm{th}(T) = s(T\gtrsim 100~\mathrm{K}) = \mathcal{H}/T$ is $\propto T^2$ (inset of Fig.~\ref{figure_2}d). The observed magnitude of $s_\mathrm{th}(T) = cs_\mathrm{theory}(T)$ agrees closely with the theoretically predicted entropy density $s_\mathrm{theory}(T)$ \cite{yudhistira2023non}, where the numerical constant $c$ lies between $0.5$ and $2$ for all devices measured. Subsequently, multiplying $\sigma$ to the fitted expression of ${\cal L}(n)$ yields $n$-dependence of $\kappa_\mathrm{e}$ as shown for three values of $T > T_\mathrm{imp}$ in Fig.~\ref{figure_3}a. Obtaining $\kappa_\mathrm{e}$ and $\sigma$ for the same $n$ ($\lesssim 10^{11}$~cm$^{-2}$), we find $\kappa_\mathrm{e}$ to {\it decrease} with {\it increasing} $\sigma$ (Fig.~\ref{figure_3}b), which cannot be explained within a Fermi liquid framework, and points towards the relativistic hydrodynamic description described by Eqn.~\ref{kappa}. 
 
Combining $\kappa_\mathrm{e}$, $\sigma$ and $s_\mathrm{th}$, each of which is obtained from either directly or through analysis of the transport and noise measurements, we then compute $\sigma_\mathrm{Q} = \kappa_\mathrm{e}\sigma/v_\mathrm{F}l_\mathrm{mr}s_\mathrm{th}$ (Details in Supplementary Section S8), and first show it as a function of $T$ for different $n$ in Fig.~\ref{figure_3}c. Beyond the inhomogeneous puddle-dominated regime ({\it i.e.} $T \lesssim 80$~K) (shaded region in Fig.~\ref{figure_3}c), it is evident that close to the charge neutrality point ($|n|\lesssim 5 \times 10^{10}$~cm$^{-2}$), $\sigma_\mathrm{Q}$ approaches $\approx 4e^2/h$, and is nearly independent of temperature. For larger $n$ ($\gtrsim 5 \times 10^{10}$~cm$^{-2}$), $\sigma_\mathrm{Q}$ is lower than this value in both the electron and the hole-doped regimes. This is further emphasized when $\sigma_\mathrm{Q}$ is plotted against $T_\mathrm{F}/T$ for different devices and temperatures ($\gtrsim T_\mathrm{imp}$), where it follows very similar trajectories that converge to $(4\pm1)\times e^2/h$ for $T_\mathrm{F}/T \ll 1$, and drops sharply for $T_\mathrm{F}/T \gg 1$ (Fig.~\ref{figure_3}d). To distinguish between $\sigma_\mathrm{min}$ and $\sigma_\mathrm{Q}$, we have compared these two parameters for all devices in Fig.~\ref{figure_3}e. $\sigma_\mathrm{min}$ is clearly non-universal and increases rapidly with disorder ($n_\mathrm{min}(0)$) by more than an order with the dashed line suggesting $\sim n^2_\mathrm{min}(0)$ dependence, whereas $\sigma_\mathrm{Q}$ varies by $\lesssim 25\%$ (shaded region). 

The device independence of $\sigma_\mathrm{Q}$ can be readily attributed to the universality of dc conductivity of hydrodynamic transport in the presence of incoherent electron-hole collisions, for example, that at superfluid-insulator transition~\cite{damle1997critical, fisher1990presence}. This quantum critical conductivity is predicted to be quantized to $4e^2\Phi/h$, where $\Phi \sim \cal{O}$[1] is a dimensionless number dependent only on the universality class of the critical point~\cite{hartnoll2007theory, damle1997critical}. The observed convergence of $\sigma_\mathrm{Q}$ to $\sim 4e^2/h$ in Fig.~\ref{figure_3}d for $T_\mathrm{F}/T\ll1$ is consistent with this scenario. As $|n|$ increases away from the Dirac point, $\sigma_\mathrm{Q}$ scales as $\sim (4e^2/h)f(T/T_\mathrm{F})$, where $f$ is a function of the dimensionless parameter $T_\mathrm{F}/T$. For $T_\mathrm{F}/T \gtrsim 1$, $f$ decreases rapidly, which is because of the suppression of the relativistic physics and quantum criticality as the system crosses over from the zero to the finite-momentum mode \cite{muller2008quantum}.

\begin{figure}[t!]	 
     \includegraphics[width=0.5\textwidth]{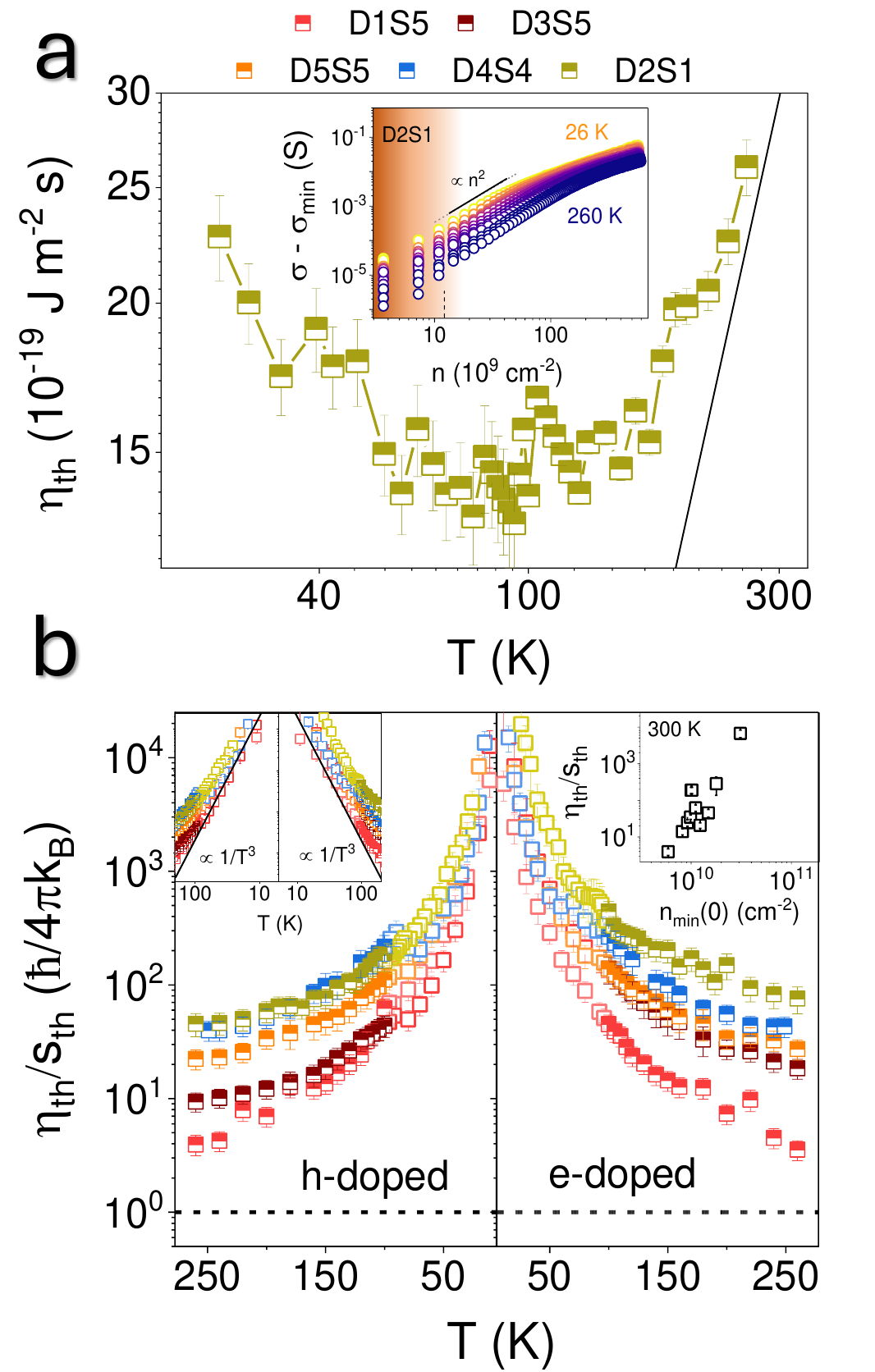}
	%\makebox[\textwidth][l]{\includegraphics[width=0.7\textwidth]{Fig_4.pdf}}
	\caption{ \textbf{Thermal viscosity in ultra-clean graphene:} (a) Thermal shear viscosity ($\eta_\mathrm{th}$) for D2S1 as a function of $T$. The solid line indicates an asymptotic $T^2$ dependence. [Inset] The quadratic density dependence of $\sigma - \sigma_\mathrm{min}$ where the inhomogeneity regime ($n < n_\mathrm{min}(0) \approx 10^{10}$~cm$^{-2}$) is indicated through shading. The upper limit of this range is marked by a dashed line, which represents $n_\mathrm{min}(0)$.(b) $\eta_\mathrm{th}/s_\mathrm{th}$ as a function of $T$ for both the electron- and hole-doped regimes, in five different devices. The dashed line indicates the holographic lower bound. The data for $T < T_\mathrm{imp}$ have been indicated using open symbols. [Left inset] $\eta_\mathrm{th}/s_\mathrm{th}$ in double logarithmic scale to indicate the power law dependence on temperature (solid lines). [Right inset] Dependence of $\eta_\mathrm{th}/s_\mathrm{th}$ on $n_\mathrm{min}(0)$ at $T = 300$~K.}
	\label{figure_4}
\end{figure}

\section{Discussion}
With the temperature as the only energy scale left, we will now estimate the effective viscosity $\eta_\mathrm{th} = {\cal H}_\mathrm{th}\tau_\mathrm{P}/2$ in the thermal regime, where we assume the current relaxation to occur by electron-hole collisions at the Planckian rate $\tau_\mathrm{P}^{-1} \approx \alpha^2k_\mathrm{B}T/\hbar$ \cite{gallagher2019quantum}. Notably, no such assumption was required in the evaluation of the quantum-critical conductivity $\sigma_\mathrm{Q}$. The hydrodynamicity of charge flow is then expected to result in a quadratic $n$-dependence of $\sigma$ (Eqn.~\ref{sigma}), which we can indeed identify in our high-quality devices between $|n|\gtrsim n_\mathrm{min}(0) \sim 10^{10}$~cm$^{-2}$, and $T_\mathrm{F} \lesssim T$, as shown for D2S1 in the inset of Fig.~\ref{figure_4}a (see Extended Data Fig.~6 for other devices). This allows us to estimate $\eta_\mathrm{th} = e^2v_\mathrm{F}W\tau_\mathrm{P}/2A_\mathrm{th}(T)$, where $A_\mathrm{th}(T)$ is the coefficient of $n^2$ in the $n$-dependence of $\sigma$. 

$\eta_\mathrm{th}$ varies nonmonotonically with a minimum around $T \sim T_\mathrm{imp}$. For $T\gtrsim T_\mathrm{imp}$, $\eta_\mathrm{th}$ increases and approaches $\sim T^2$ dependence at high $T$ \cite{yudhistira2023non} (Fig.~\ref{figure_4}a, details in Supplementary Section S9). Finally, with $\eta_\mathrm{th}$ estimated from the $\sigma$ as above and $s_\mathrm{th}$ obtained from thermal conductivity (dashed line in Fig.~\ref{figure_2}d, inset), we have computed the ratio $\eta_\mathrm{th}/s_\mathrm{th}$ normalized by its universal limit $\hbar/4\pi k_\mathrm{B}$ \cite{muller2009graphene}. Within the holographic description, the normalised $\eta_\mathrm{th}/s_\mathrm{th}$ is expected to approach unity for minimally dissipative flow of strongly interacting quantum liquids, limited only by the Heisenberg's uncertainty principle \cite{kovtun2005viscosity}. We have plotted the normalised $\eta_\mathrm{th}/s_\mathrm{th}$ as a function of $T$ for both electron and hole-doped regimes for several devices in Fig.~\ref{figure_4}b. At $T<T_\mathrm{imp}~\sim 100$~K, the variation in $\eta_\mathrm{th}/s_\mathrm{th} \sim 1/T^3$, represented by the open symbols and solid lines in the left inset of Fig.~\ref{figure_4}b, is consistent with that expected from a degenerate Fermi liquid \cite{muller2009graphene}, where the chemical potential close to the Dirac point is pinned to the inhomogeneity scale (Fig.~\ref{figure_2}e). For $T \gg T_\mathrm{imp}$, however, $\eta_\mathrm{th}/s_\mathrm{th}$ tends to saturate, but the saturation occurs to lower values when the channel disorder is decreased, and for the cleanest channel D1S5 (see Table~1), $\eta_\mathrm{th}/s_\mathrm{th}$ approaches $\hbar/4\pi k_\mathrm{B}$ within a factor of 4. The sensitivity of the limiting $\eta_\mathrm{th}/s_\mathrm{th}$ to disorder, quantified by $n_\mathrm{min}(0)$, in Fig.~\ref{figure_4}b (right inset), is quite striking, and could be due to increase in the shear viscosity in the presence of disorder scattering in Dirac fluids \cite{chen2022viscosity}.

In conclusion, we report signatures of viscous electron flow and quantum critical transport in extremely clean monolayer graphene devices with charge inhomogeneity scale as low as $\sim 5\times 10^9$~cm$^{-2}$. At the charge neutrality point, we find the electronic thermal conductivity to exceed that expected from the WF law by nearly 300 times at low temperatures. Combining electrical and thermal transport measurements, we could evaluate the intrinsic dc quantum critical conductivity $\sigma_\mathrm{Q} \approx 4e^2/h$, with device-to-device variation $\lesssim \pm25\%$.

Considering a Planckian scattering-limited electronic viscosity, we also demonstrate that the ratio of viscosity to thermal entropy density $\rightarrow \hbar/4\pi k_\mathrm{B}$, the holographic limit, within about a factor of four in the cleanest devices at room temperature. Apart from unifying some of the deep conceptual frameworks in physics,  our results will impact the analysis and interpretation of both dc charge and heat transport in monolayer graphene, as the quality of such devices continues to improve. 

\section{Methods}

\subsection{Fabrication of hBN-encapsulated graphene heterostructures}

All the devices presented in this work have been assembled using a manually operated transfer setup. The graphene (Gr) and hexagonal boron nitride (hBN) flakes were mechanically exfoliated from bulk graphite (obtained from KishGraphite) and bulk hBN crystals (obtained from NIMS, Japan) respectively, using a combination of both scotch-tape and hot exfoliation. Subsequently, the thickness of the flakes was confirmed using atomic force microscopy (AFM) and Raman spectroscopy. Further, the cleanliness of the individual flakes was qualitatively characterised using the surface roughness of flakes obtained from AFM line scans and the peak line widths obtained from Raman spectroscopy (Supplementary Section S2). These flakes were thereafter stacked into a van der Waals (vdW) heterostructure on a Si/SiO$_2$ substrate using a dry-transfer technique that involves coating a PPC (Polypropylene Carbonate) film on a hemispherical PDMS (Polydimethylsiloxane) drop. The transferred vdW heterostructure was etched into a rectangular shape using e-beam lithography, followed by reactive ion etching using a mixture of O$_2$ and CHF$_3$ gases. Finally, electrical contacts were deposited on the heterostructure via thermal evaporation of Cr ($5$~nm) and Au ($60$~nm).

\subsection{Transport measurements}

The electrical and thermal transport characterisation of the devices was carried out in a home-made He-4 Cryostat and in the s200 4K Cryostation (Montana Instruments), respectively. The temperature-dependent transfer characteristics were measured using a SR830 Lock-in Amplifier and a Keithley 2400 Source Meter. The differential conductance measurements were carried out using an AC-DC mixer circuit, where the AC signal was sourced and measured using a Lock-in Amplifier, while the DC signal was sourced from a Keithley 2400 Source Meter and measured using a Keithley 2182A Nanovoltmeter. The thermal transport measurements were carried out using a high-frequency compatible Johnson noise thermometry setup, whose schematic has been detailed in Supplementary Section S4. 

\subsection{Solution of the heat diffusion equation}

The DC electric field applied across the contacts of the device leads to the formation of a thermal gradient across the device. The spatial distribution of the thermal gradient is ideally governed by the heat diffusion equation. However, under experimental conditions, it can be reduced to the 1D Fourier's equation, which is given by
\begin{equation}
    q = - \kappa_\mathrm{e} \nabla T_\mathrm{e}(x)
\end{equation}
where $q$ is the rate of heat flow into the channel and $T_\mathrm{e}(x)$ is the electron temperature at any point x along the direction of current flow.

For our graphene device, the rate of power influx is equal to the power dissipated at the contacts. 
\begin{equation}
    q = J W E L
\end{equation}
where $W$ and $L$ are, respectively, the width and length of the channel.

If we assume the thermal gradient to primarily be in the direction of the applied electric field, the above equation can be solved to get the following solution:
\[ T_\mathrm{e}(x) = \begin{cases*}
                    \phantom{-} \sqrt{T_\mathrm{c}^2 + \dfrac{2L}{\cal{L}}E^2 x} & for $0 \leq x \leq L/2$  \\
                     \phantom{-}\sqrt{T_\mathrm{c}^2 + \dfrac{2L}{\cal{L}}E^2(L-x)} & for $L/2 \leq x \leq L$
                 \end{cases*} \]

where $T_\mathrm{c}$ is the temperature of the metallic electrode.

The measured thermal noise gives us an average electron temperature ($T_\mathrm{e}$), integrated over the entire surface area of the device and it is given by
\begin{equation}
    T_\mathrm{e} = \dfrac{2\cal L}{3L^2E^2}\left[\left(T_\mathrm{c}^2 + \dfrac{L^2}{\cal L} E^2 \right)^{3/2} - \left(T_\mathrm{c}^2\right)^{3/2}\right]
    \label{Te}
\end{equation}

Detailed derivation of this calculation has been presented in Supplementary Section S10.

\section{Data Availability}
All data files are available from the corresponding author upon request.

\section{Code Availability}
The code used for theoretical simulations present in the Supplementary Information is available in the linked \href{https://github.com/NisargChadha/Quantum_Critical_Fluid/blob/main/Mathiessen.py}{Github} repository.

\section{Acknowledgment}
The authors gratefully acknowledge the usage of the MNCF and NNFC facilities at CeNSE, IISc. The authors would also like to acknowledge fruitful discussions with D. Sen, A. Lucas, S. Sachdev, A. Hui, B. Skinner, N. Trivedi, M. Randeria, B. Dora, R. Moessner, A. Green and S. Sondhi. A.G. acknowledges financial support from a project under NanoMission, Department of Science and Technology, India and J. C. Bose Fellowship. P.P. and Ak.G. thank the Ministry of Education, Govt. of India for the Prime Minister's Research Fellowship (PMRF). K.W. and T.T. acknowledge support from the JSPS KAKENHI (Grant Numbers 21H05233 and 23H02052) and World Premier International Research Center Initiative (WPI), MEXT, Japan.

\section{Ethics Declaration}
The authors declare no competing interests.

\bibliography{bibliography}

% Redefine figure captions
\renewcommand{\figurename}{Extended Data Figure}

\makeatletter
\@fpsep\textheight
\makeatother

% Reset the figure counter to 1
\setcounter{figure}{0}

\begin{figure*}
	\includegraphics[width=1.0\textwidth]{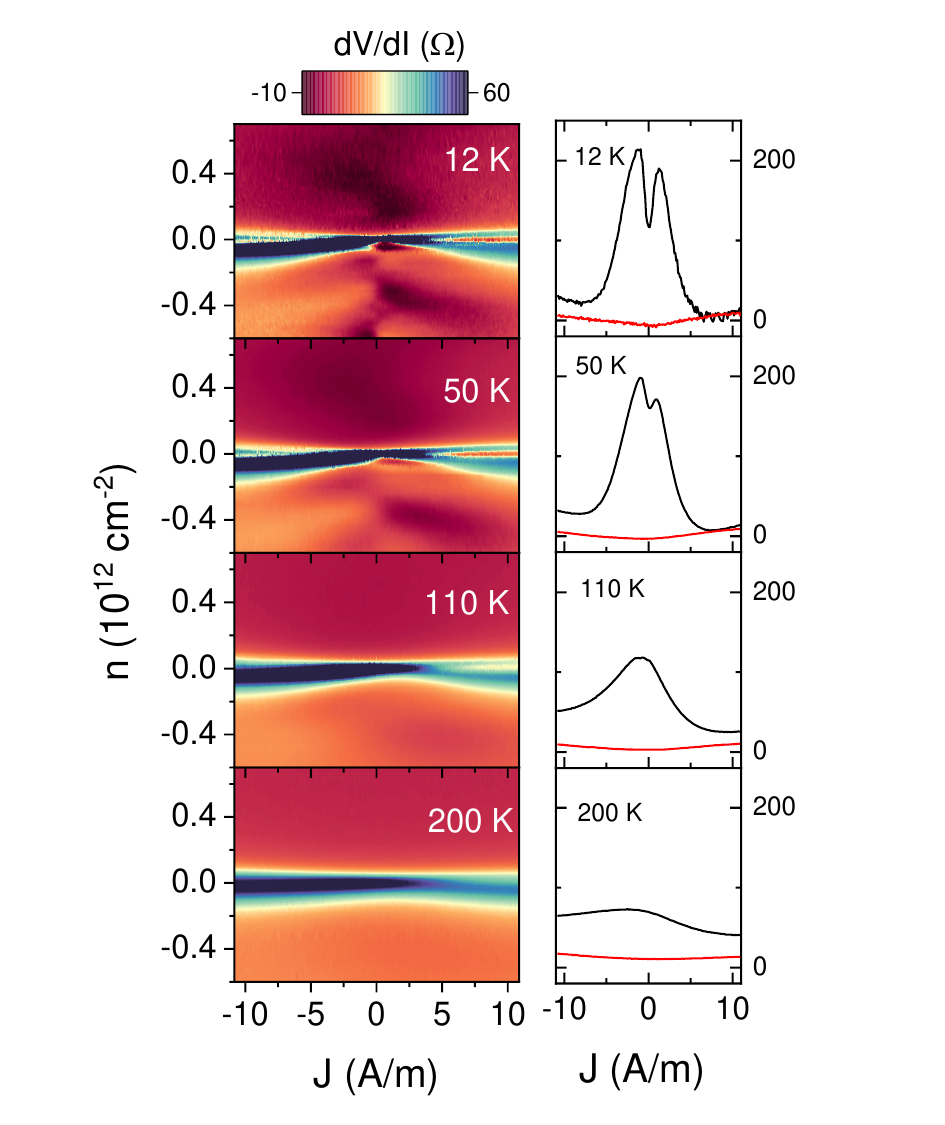}
        \centering
	\caption{ \textbf{Differential resistance measurements showing crossover from Knudsen to Poiseuille regime:} (Left side) Colour plot showing the variation of $dV/dI$ for D3S5 as a function of $n$ and applied electrical current density ($J$) at four different $T$. (Right side) Line plots, obtained from the respective colour plots on the left, depicting the variation of $dV/dI$ as a function of applied electrical current density ($J$) for two distinct number densities: $n = 10^{10}$~cm$^{-2}$ (marked in black) and $n = 10^{11}$~cm$^{-2}$ (marked in red). }
	\label{Extended_Data_Fig_1}
\end{figure*}

\begin{figure*}
	\includegraphics[width=0.8\textwidth]{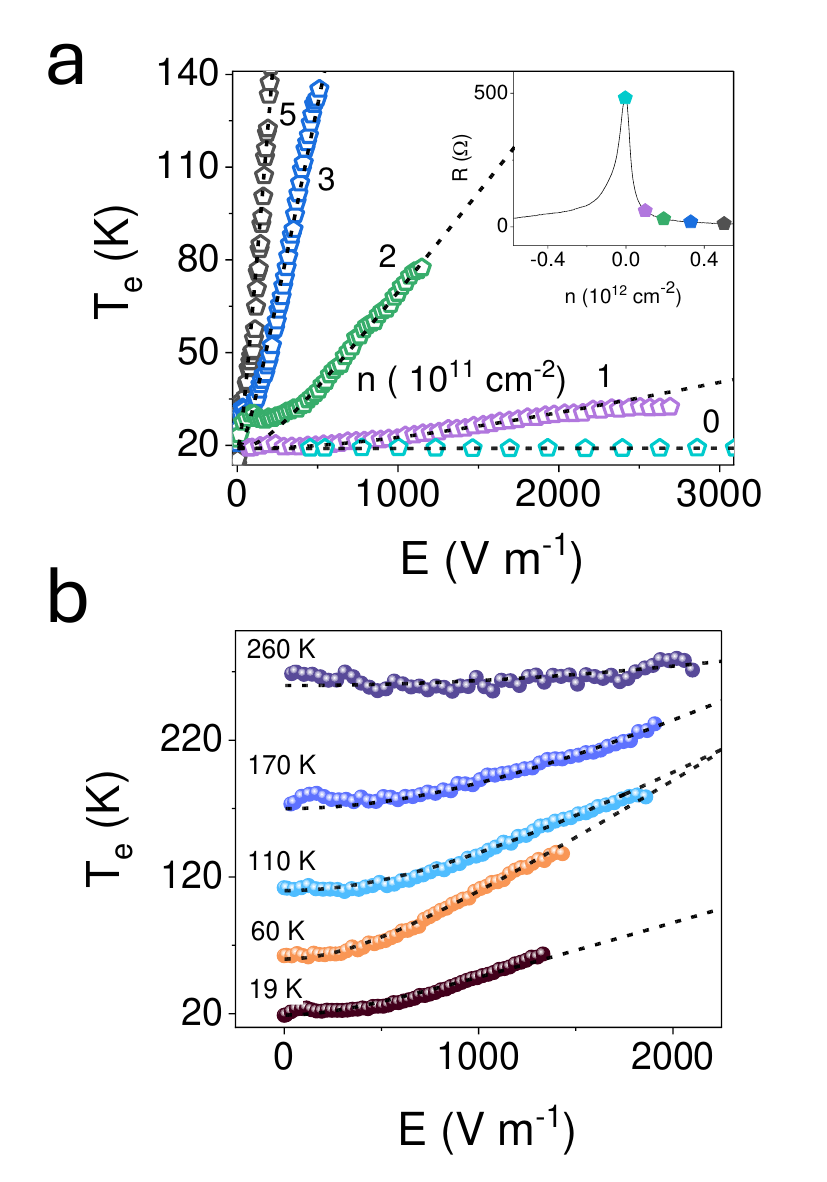}
	\caption{ \textbf{Electronic temperature across the graphene channel at low electric fields:} (a)  $T_\mathrm{e}$ in D2S1 as a function of the applied electric field ($E$) at different electron densities, recorded at $T = 19$~K. The dashed lines are theoretically fitted curves for the experimental data, following Eqn.~\ref{Te}. [Inset] Transfer characteristics of the device D2S1 at $19$~K, with colored dots indicating the resistance at the different $n$ at which $T_\mathrm{e}$ vs $E$ data has been recorded. (b)  $T_\mathrm{e}$ as a function of the applied electric field ($E$) at different temperatures, for $n = 5 \times 10^{11}$~cm$^{-2}$.}
	\label{Extended_Data_Fig_2}
\end{figure*}

\begin{figure*}
	\includegraphics[width=1.0\textwidth]{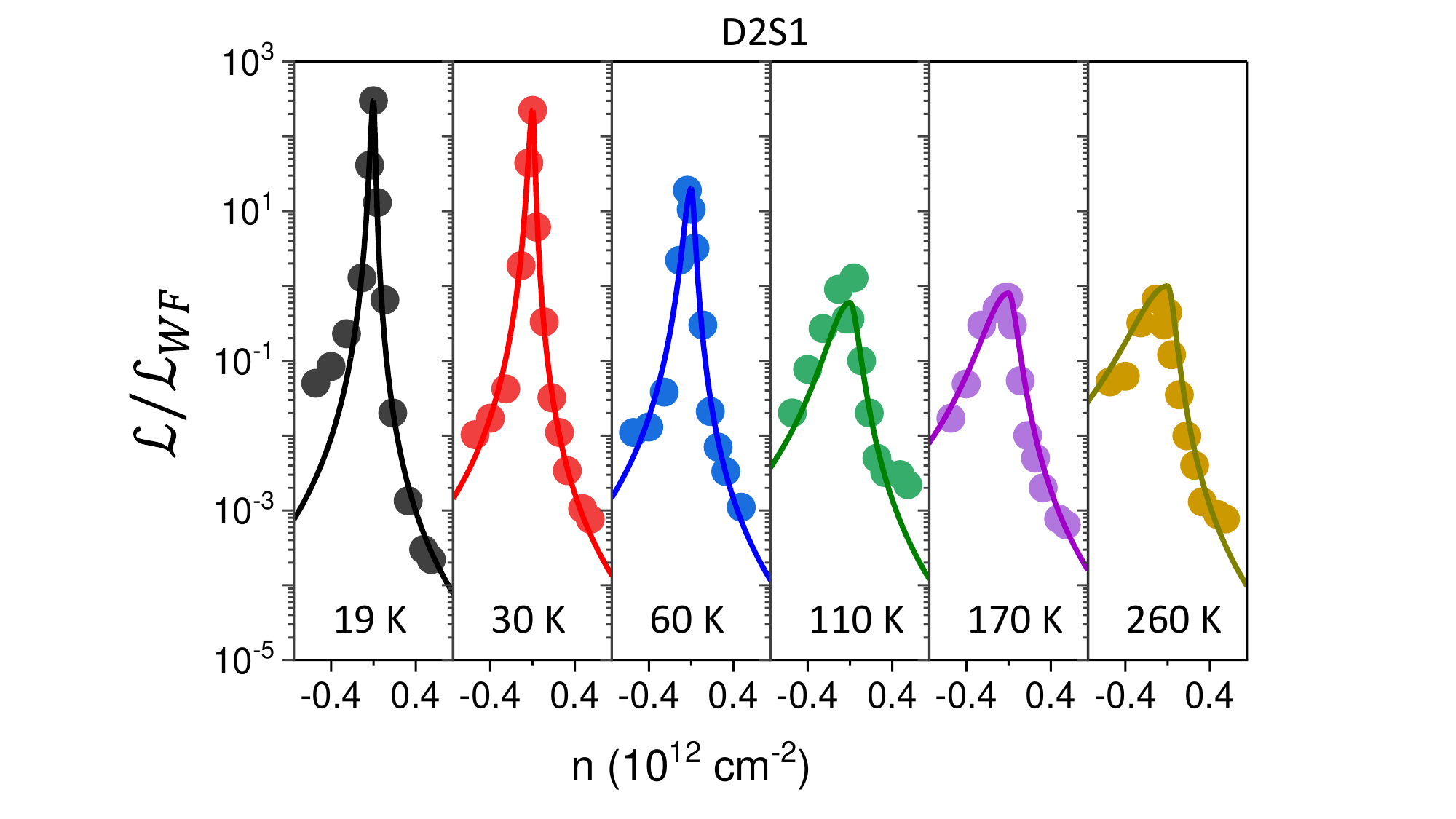}
	\caption{\textbf{Violation of WF Law in D2S1 at six different temperatures:} Normalised Lorentz ratio (${\cal L}/{\cal L}_{WF}$) for D2S1, as a function of $n$ for $T = 19$~K, $30$~K, $60$~K, $110$~K, $170$~K and $260$~K. The solid lines are theoretical fits for the experimental data, as per Eqn.~\ref{L}.} 
	\label{Extended_Data_Fig_3}
\end{figure*}

\begin{figure*}
	\includegraphics[width=0.65\textwidth]{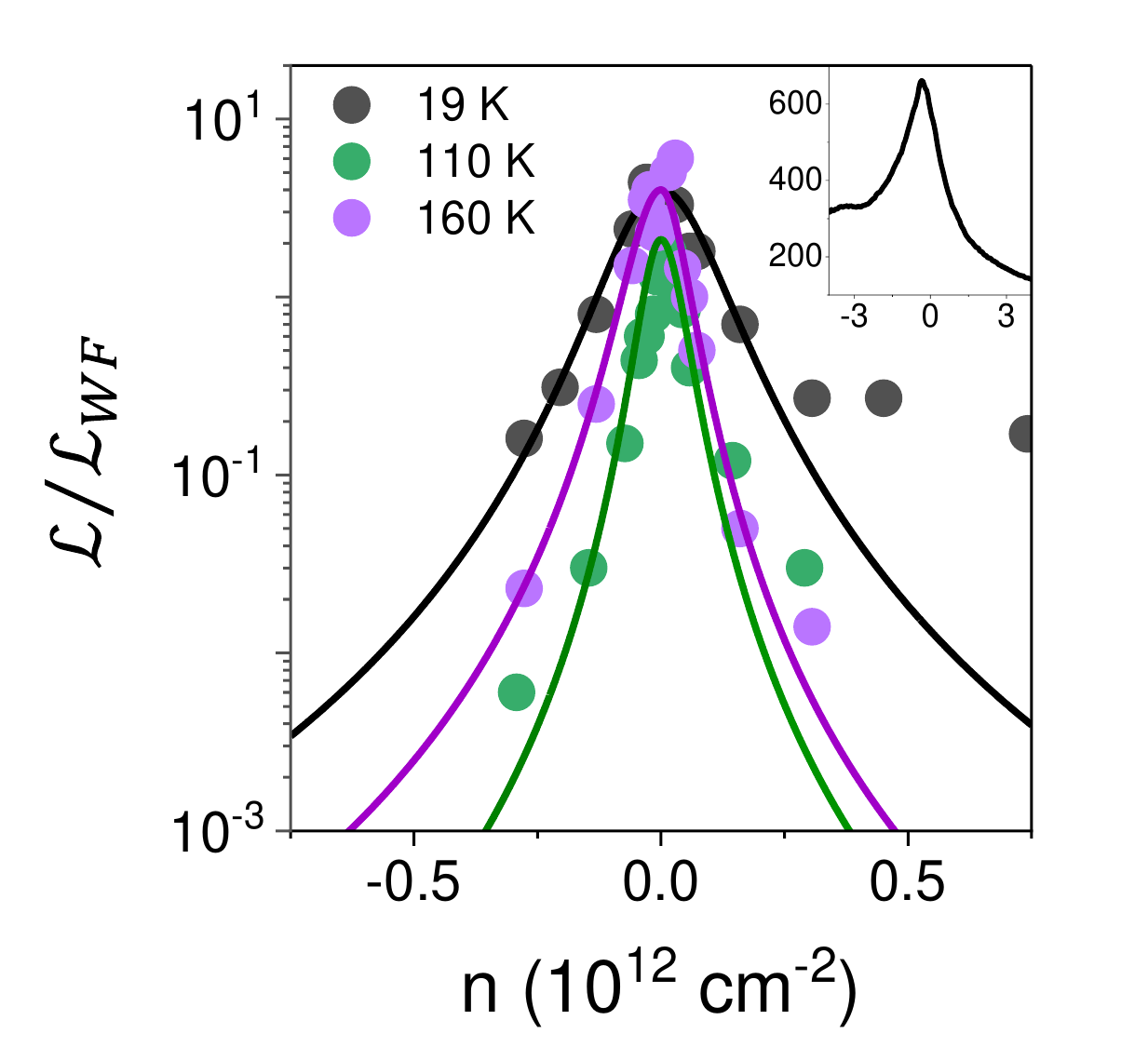}
	\caption{ \textbf{Carrier density- and temperature-dependence of the Lorentz number in D1S1: } Normalised Lorentz ratio (${\cal L}/{\cal L}_{WF}$) for D1S1, as a function of $n$ for three different $T$. The solid lines are theoretical fits for the experimental data, as per Eqn.~\ref{L}. Inset shows the transfer characteristics of D1S1 at $T = 40$~K. The y-axis shows the electrical resistance in ohms while the x-axis shows the applied gate voltage $V_\mathrm{g}$ in volts. Further electrical characterisation of D1S1 indicating its intrinsic charge inhomogeneity and mobility are mentioned in Table~I in Supplementary Information.}
	\label{Extended_Data_Fig_4}
\end{figure*}

\begin{figure*}
	\includegraphics[width=0.6\textwidth]{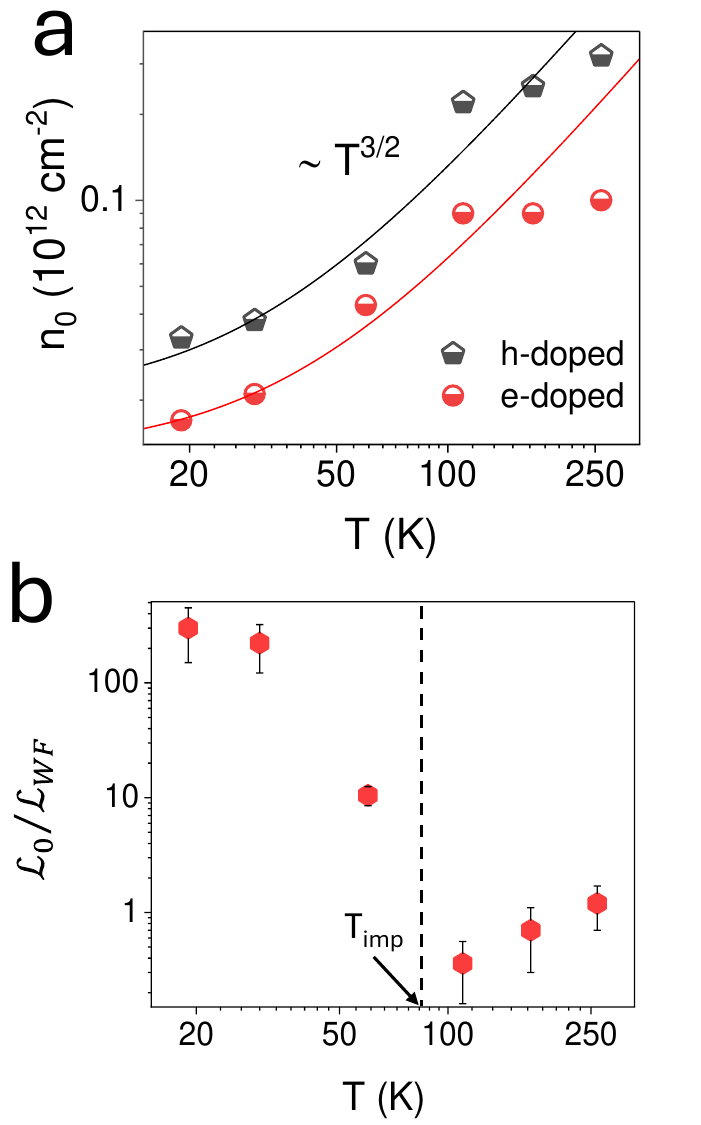}
	\caption{ \textbf{Characteristic scales of Lorentz number variation with density and temperature in D2S1:}  (a)  The characteristic density scale $n_0$ as a function of T, for the electron- and hole-doped ranges of D2S1. The solid lines show a $T^{3/2}$ fit of the experimental data points. (b) Normalised Lorentz ratio at the Dirac point (${\cal L}_{0}/{\cal L}_{WF}$) of D2S1, as a function of T. The dashed line indicates the $T_\mathrm{imp}$ for D2S1.}
	\label{Extended_Data_Fig_5}
\end{figure*}

\begin{figure*}
	\includegraphics[width=1.0\textwidth]{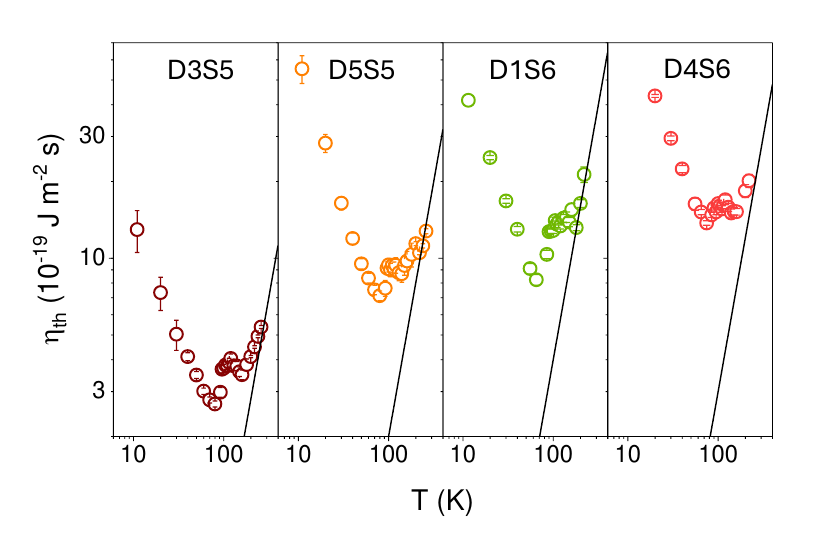}
	\caption{\textbf{Non-monotonic temperature dependence of electron viscosity for additional devices:} $\eta_\mathrm{th}$ vs $T$ for four devices D3S5, D5S5, D1S6 and D4S6. The non-monotonic trend is consistent across all the devices and approaches $\sim T^2$ for $T > T_\mathrm{imp}$. The black solid lines in each panel represent $T^2$ dependence.  }
	\label{Extended_Data_Fig_6}
\end{figure*}

\end{document}